\documentclass[conference]{IEEEtran}

\usepackage[numbers]{natbib}
\usepackage{graphicx}
\usepackage{textpos}
\usepackage{float}
\usepackage{subfigure, multirow} 
\usepackage{amsmath, amsthm, amssymb, algorithm2e, subfigure, balance}
\usepackage{multicol}
\usepackage{algpseudocode}
\usepackage{url}
\usepackage{listings}
\usepackage{soul}

\usepackage{fancyhdr}

\usepackage{hyperref}

\usepackage{amsfonts}
\usepackage{amssymb}
\usepackage{tabularx}
\usepackage{amsmath}
\usepackage{tikz}
\usepackage{graphicx}
\usepackage{caption}

\usetikzlibrary{arrows}
\tikzset{edge/.style = {->,> = latex'}}

\makeatletter
\newsavebox\myboxA
\newsavebox\myboxB
\newlength\mylenA

\newcommand*\xoverline[2][0.75]{%
	\sbox{\myboxA}{$\m@th#2$}%
	\setbox\myboxB\null
	\ht\myboxB=\ht\myboxA%
	\dp\myboxB=\dp\myboxA%
	\wd\myboxB=#1\wd\myboxA
	\sbox\myboxB{$\m@th\overline{\copy\myboxB}$}
	\setlength\mylenA{\the\wd\myboxA}
	\addtolength\mylenA{-\the\wd\myboxB}%
	\ifdim\wd\myboxB<\wd\myboxA%
	\rlap{\hskip 0.5\mylenA\usebox\myboxB}{\usebox\myboxA}%
	\else
	\hskip -0.5\mylenA\rlap{\usebox\myboxA}{\hskip 0.5\mylenA\usebox\myboxB}%
	\fi}
\makeatother

\begin{document}

\title{On the Generalizability of Linear and Non-Linear Region of Interest-Based Multivariate Regression Models for fMRI Data}

\author{
	\IEEEauthorblockN{Ethan C. Jackson}
	\IEEEauthorblockA{Computer Science Department\\ University of Western Ontario\\London, Ontario, Canada N6A 3K7\\ ejacks42@uwo.ca}

	\and
	
	\IEEEauthorblockN{James Alexander Hughes}
	\IEEEauthorblockA{Computer Science Department\\ University of Western Ontario\\London, Ontario, Canada N6A 3K7\\ jhughe54@uwo.ca}
	
	\and
	
	\IEEEauthorblockN{Mark Daley}
	\IEEEauthorblockA{Computer Science Department\\ University of Western Ontario\\London, Ontario, Canada N6A 3K7\\ mdaley2@uwo.ca}
}

\maketitle

\begin{abstract}
In contrast to conventional, univariate analysis, various types of multivariate analysis have been applied to functional magnetic resonance imaging (fMRI) data. In this paper, we compare two contemporary approaches for multivariate regression on task-based fMRI data: linear regression with ridge regularization and non-linear symbolic regression using genetic programming. The data for this project is representative of a contemporary fMRI experimental design for visual stimuli. Linear and non-linear models were generated for 10 subjects, with another 4 withheld for validation. Model quality is evaluated by comparing $R$ scores (Pearson product-moment correlation) in various contexts, including single run self-fit, within-subject generalization, and between-subject generalization. Propensity for modelling strategies to overfit is estimated using a separate resting state scan. Results suggest that neither method is objectively or inherently better than the other.
\end{abstract}


\section{Introduction}

fMRI is a non-invasive neuroimaging technique used in research and clinical applications to record relative changes in metabolic rate associated with neural activity via the blood oxygenation level dependent (BOLD) signal \cite{disbrow2000functional}. This signal measures the relative levels of oxygenated and deoxygenated blood throughout the whole brain, and is a known metabolic proxy for neural activity.

The neuroimaging field has been very successful in using linear statistical methods for analysis. The general linear model (GLM), in particular, has been fruitfully applied to identify relationships between experimental conditions and localized BOLD signal. To fit a GLM, neuroscientists create a predictive time series model (predictor) for each experimental condition of interest. Usually, these predictors are square-wave time series corresponding to stimulus onset and duration convolved with a model function for haemodynamic response (HR). For each voxel (a volumetric pixel) or aggregate region of interest (ROI), a multiple linear regression is performed to compute a linear combination of predictors that optimally models the recorded BOLD signal. A voxel or ROI's relationship to an experimental condition can be inferred by comparing the relative weights (scaling factors) for each of the predictors. A variety of statistical methods can be applied to perform comparisons over all voxels or ROIs, though paired t-tests are commonly used.

In contrast to conventional univariate analysis, multivariate relational modelling has also been developed. For example, in Hughes and Daley's work on relational modelling between ROI average time series (\cite{hughes2016finding}, \cite{hughes2017searching}), average BOLD in one ROI (dependent variable) is modelled as a non-linear function of BOLD signal in other ROIs (independent variables). In order to make a more direct comparison to conventional linear regression and the GLM, the dependent regression variable can be replaced with a hypothesized, stimulus-driven HR function. In this configuration, regression models HR as a spatially distributed combination of BOLD signal --- an approach that has applications in \textit{multivoxel pattern analysis} \cite{norman2006beyond}. 

Naturally one might ask whether conventional, linear regression or much more elaborate non-linear symbolic regression is better suited for this type of modelling. This paper aims to answer that question by comparing the generalizability of models generated using each method. We assess generalizability based on within- and between-subject analytical experiments, and we further estimate each method's lower bound for propensity to overfit using an additional resting state scan. The fMRI data consists of scans from 15 subjects in total --- a decidedly small but generally acceptable sample size in the neuroimaging community.

In the following sections, we first give details about the data used in our experiments. Next, we give an overview of the experiments themselves, followed by sections explaining details about linear and non-linear models, respectively. Then we present results, discussion, and finally avenues for future work.

\section{Data and Preprocessing}\label{Data}

For all experimental runs, we used a 10-subject subset of fMRI data collected for didactic purposes. Validation uses four additional experimental scans and one additional resting state scan.

All functional scans were collected using the same scanner and parameters: Siemens Magnetom Prisma 3T, 2.5-mm isotropic resolution with matrix size of 84$\times$84 over 52 slices, TR = 1000ms, multi-band gradient-echo echoplanar pulse sequence. Anatomical scans were collected using a matrix size of 248$\times$256 over 176 1mm slices. More information about the scanner and facilities can be found at \url{http://cfmm.robarts.ca/tools/3t-mri/}.

The experimental stimuli used were designed to localize brain regions selective to three categories of visual stimuli: faces, hands, and bodies. Scrambled versions of the images formed a fourth stimulus category.  Participants were instructed to maintain their gaze on a central fixation point, which was also presented on a blank screen during baseline periods.

Images were presented to subjects according to a conventional block design, with a block duration of 16s. Four cycles of four blocks (faces, hands, bodies, scrambled images) were presented in each run.  Baseline blocks occurred at the beginning and end of each run and between cycles. The total run duration was 336 s (21 blocks x 16 s/block). Each stimulus block consisted of 16 stimuli, each presented for 0.8 s with a 0.2 s intertrial interval.  The order of blocks within cycles was balanced such that for a run, each category was presented once in each of the first, second, third and fourth part of the cycle. Two runs with two different orders were collected. Participants monitored the stream of visual stimuli for repetitions (a “one-back task) to maintain attention.

Data were preprocessed in Brain Voyager 20.6 commercial software  (\cite{goebel2006analysis}, \cite{formisano2006fundamentals}). Preprocessing consisted of functional-anatomical alignment, three-dimensional motion correction, temporal high-pass filtering (X cycles/run), and transformation into 2-mm isotropic resolution, warping into stereotaxic space using the MNI-152 template. Data were loaded into Python using a Brain Voyager-compatible fork of NiBabel  \cite{nibabel-brainvoyager}.

\begin{figure}
	\centering
	\includegraphics[width=1.0\linewidth]{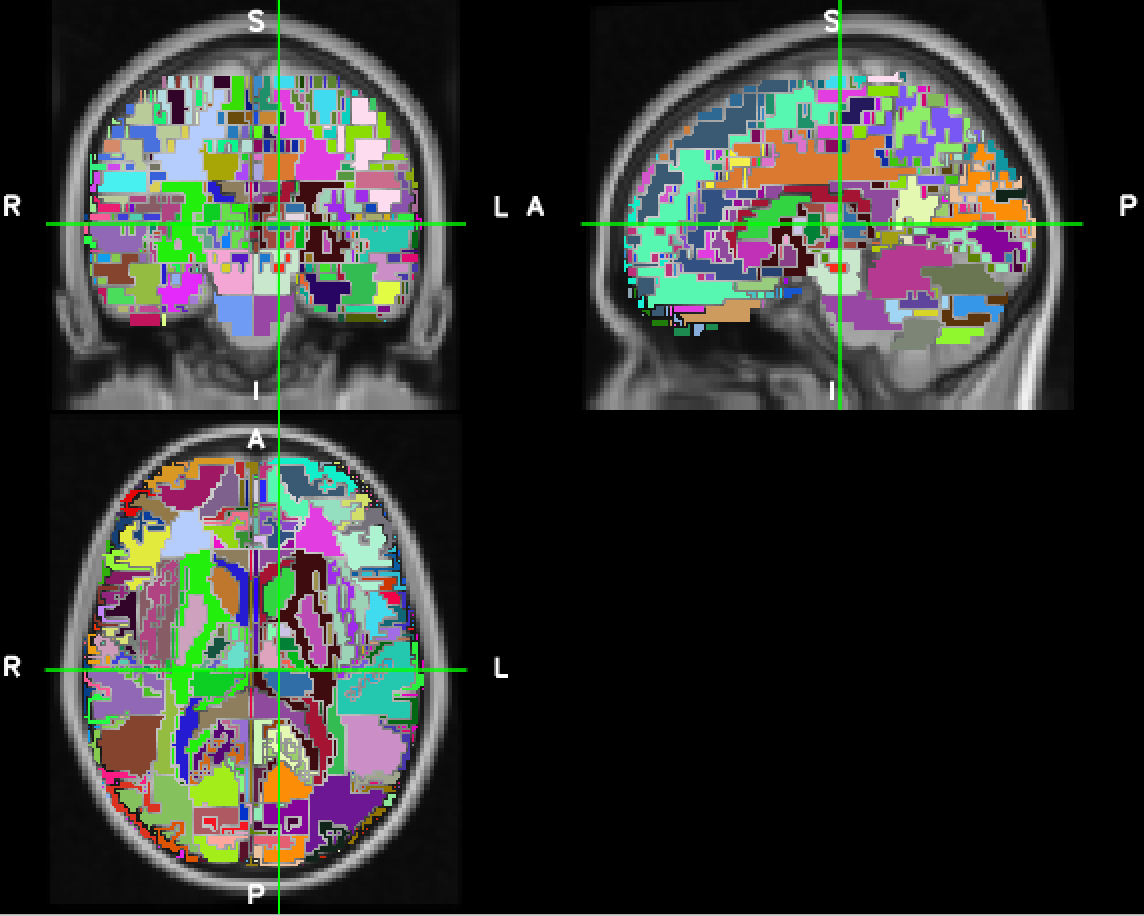}
	\caption{Talairach standard atlas used to partition the whole brain into 986 ROIs. Statistical transformations on smaller ROIs are more likely to capture highly stimulus selective metabolic responses.}
	\label{fig:talairach}
\end{figure}
\begin{figure}
	\centering
	\includegraphics[width=1.0\linewidth]{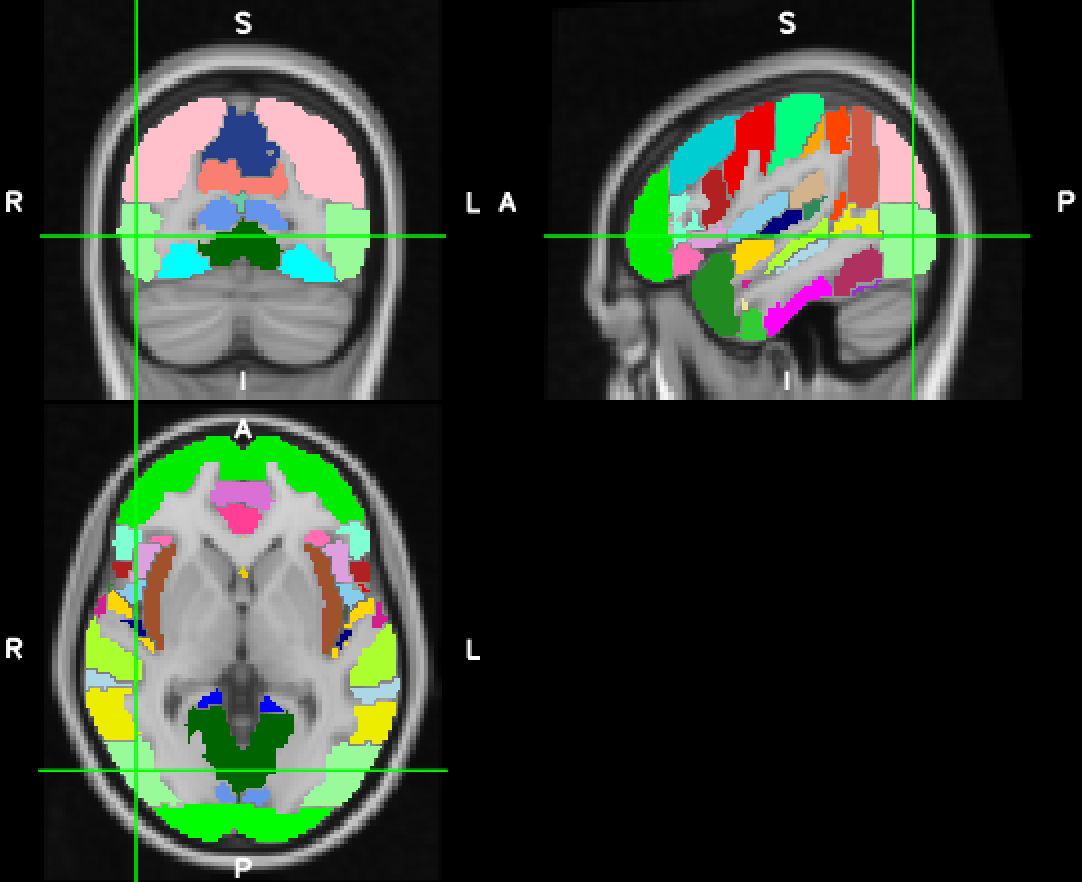}
	\caption{Harvard-Oxford atlas used to partition the cortical brain into 48 ROIs. Large ROIs are less sensitive than smaller ones to partitioning error due to anatomical differences between subjects.}
	\label{fig:harvardoxford}
\end{figure}

For analysis, we considered two types partitioning. Though various transformations could be applied to yield an aggregate time series from each ROI, we chose to use the average activation of the whole ROI. This is the same approach used in \cite{hughes2016finding} and \cite{hughes2017searching}. Many methods for defining partition boundaries exist. Conventional partitioning of cortex, for example, is performed using static maps (atlases) on a reference brain. In our experiments, we used two atlases: the Harvard-Oxford Cortical Structural (H-O) atlas, and the Talairach Daemon Labels (TalROI) atlas. Both atlases are provided with FSL \cite{jenkinson2012fsl}, and are visualized by Figures \ref{fig:harvardoxford} and \ref{fig:talairach}, respectively. The H-O atlas defines 48 ROIs, while TalROI defines 986. To avoid overfitting in linear regression and unreasonable computational complexity for non-linear regression, an additional preprocessing step was applied to TalROI.

\begin{figure}
	\centering
	\includegraphics[width=0.9\linewidth]{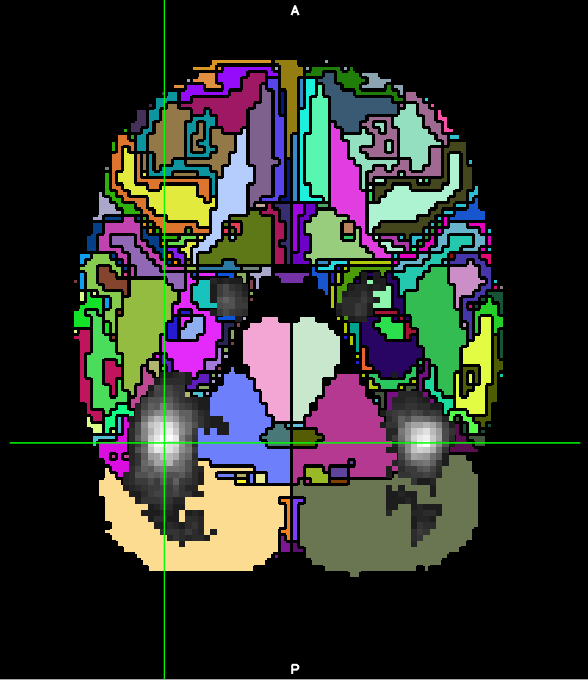}
	\caption{Two dimensional slice of Neurosynth mask for FFA overlaid on TalROI atlas. Areas coloured in grey scale indicate voxels in MNI152 space that are associated with FFA according to the Neurosynth corpus. Higher intensity (brightness) indicates stronger confidence. MNI152 coordinates $= (41.7, -51.6, -20.9)$}
	\label{fig:neurosynth}
\end{figure}

Since TalROI defines 986 ROIs and because we only used this atlas in experiments on response to face stimuli, we used a Neurosynth mask to eliminate ROIs not associated with face selectivity in its body of literature. Neurosynth~\cite{yarkoni2011large} is a web tool that uses natural language processing to identify associations between brain areas and cognitive function based on a large corpus of neuroimaging research papers. A Neurosynth search for `\textit{face}' or `\textit{FFA}' (fusiform face area) yields a mask that can be overlaid onto a functional scan in MNI space. The mask used in this work is visualized by Figure \ref{fig:neurosynth}. Application of this mask not only reduces the number of ROIs in TalROI to 50, but focuses model generation on ROIs known to demonstrate selectivity for faces.

Altogether, the fMRI data consists of 20 experimental runs, 10 each for two orders of the block-designed stimulus presentation scheme labelled `Loc1' and `Loc2' (for Localisers 1 and 2). Additionally, a resting state scan (stimulus- and task-free, eyes closed, instructed not to fixate) was collected from one independent subject. Each run consists of 340 time points. Each of the two atlases was used to partition and aggregate the each run into ROI averages. Thus, we have two transformations (H-O and TalROI) for each of the 20 experimental runs and one resting state scan. All data are stored as CSV files with the number of rows and columns equal to the number of time points and ROIs, respectively. These files, in addition to all Python code, masks, and atlases, are available as part of the digital appendix \cite{digital_appendix}.

\section{Models}
Before describing the experiments themselves, we first describe how linear and non-linear models were constructed for analysis. 

\subsection{Linear Models}

Linear models were all constructed using linear regression with ridge or $L_2$ regularization. This method computes the linear combination of ROI time series that optimally models a hypothesized HR using ordinary least squares estimation. Thus for each run, linear models are composed of $n$ scaling factors plus one constant term, where $n$ is equal to the number of ROIs used for partitioning. Results based on linear methods are labelled `L'.

\subsection{Non-Linear Models}

Non-linear models were constructed using Hughes and Daley's GP-based system designed for symbolic regression. The number of degrees of freedom is exponentially proportional to the size of the GP language and to the number of ROIs. We used two methods for assessing the quality of non-linear models.

First, to obtain one model per run (fMRI scan, transformed) in an unbiased manner, we computed 100 independent GP runs for 10,000,000 generations each. Of the 100 candidate models, we took the model of best fit as estimated by mean squared error at the end of evolution. Thus for each run, non-linear models are expressed as a non-linear function of $n$ variables, where $n$ is again equal to the number of ROIs used for partitioning. It is important to note that the resulting non-linear models are temporally independent: the non-linear functions must be applied independently at each time point in order to reconstruct a time series. Results using this method are labelled `N-L'. The GP language or set of basis functions, settings, and parameters are summarized by Table \ref{tab:GPsettings}.

The second method is the same as the first, except it considers all 100 candidate models for a run produced by GP. This allows us to plot and analyze a distribution of $R$ scores (rather than a single value) to better understand whether GP can produce better models than linear regression. Using this method, we can search a \textit{distribution of candidate models}, and select the single, best-scoring non-linear model. Though heavily biased, this analysis is useful to evaluate \textit{potential} non-linear model quality. For example in between-subject experiments, we could select the model for subject A that generalizes best to subject B. Results using this method are labelled `N-L (best)'.

\begin{table}
	\small
	\centering
	\caption{Parameter settings for GP. The last four parameters are specific to the improvements made in Hughes and Daley's GP system. Improvements include the use of an acyclic graph representation, fitness predictors, and island-model migration strategies.}
	\label{tab:GPsettings}
	\begin{tabular}{|c||c|}
		\hline Elitism 				& 1 \\ 
		\hline Population 			& 101/subpopulation (707 total) \\ 
		\hline Subpopulations 		& 7 \\ 
		\hline Migrations 			& 1,000 \\ 
		\hline Generations 			& 1,000 \textit{per} migration (10,000,000 total) \\ 
		\hline Crossover 			& 80\% \\ 
		\hline Mutation 			& 10\% (x2 chances) \\ 
		\hline Fitness Metric 		& Mean Squared Error: {\tiny $\frac{1}{n}\sum_{i=1}^{n}(\hat{Y_{i}} - Y_{i})^{2}$}\\
		\hline Language & $+$, $-$ ,$*$ , $/$, $exp$, $abs$, $sin$, $cos$, $tan$ \\ 
		\hline 
		\hline Trainers 			& 8 \\ 
		\hline Predictors  			& 20 \\ 
		\hline Predictor Pop. Size 	& 20\% of Dataset \\ 
		\hline Max \# Graph Nodes 	& 140 \\
		\hline
	\end{tabular} 
\end{table}

\section{Analytical Experiments}

\begin{figure}
	\centering
	\includegraphics[width=0.9\linewidth]{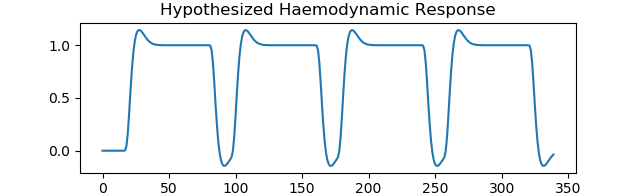}
	\caption{Hypothesized haemodynamic response to all-stimuli in arbitrary units over time in seconds for `Loc1'. This curve is obtained by taking the sum of the four stimulus-specific hypothesized HRs.}
	\label{fig:allpred}
\end{figure}

\begin{figure}
	\centering
	\includegraphics[width=0.9\linewidth]{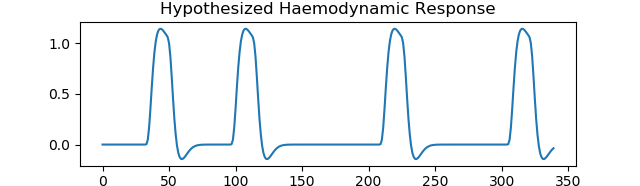}
	\caption{Hypothesized haemodynamic response to face stimuli in arbitrary units over time in seconds for `Loc1'.}
	\label{fig:facepred}
\end{figure}

To compare the quality of models obtained from multivariate linear regression to symbolic non-linear regression, and to estimate overfitting, we designed the following experiments. Experiments are categorized as either \textit{all-stims} or \textit{faces}. For the former, models fit data to a hypothesized HR to any stimulus category, as visualized by Figure \ref{fig:allpred}. For the latter, models fit data to a hypothesized HR for the \textit{faces} category of stimuli only, as visualized by Figure \ref{fig:facepred}. In practice, the set of experiments for the latter is much more interesting since it will gauge stimulus specific selectivity. The \textit{all-stims} and \textit{faces} experiments exclusively use the H-O and TalROI atlases, respectively. 

For all experiments, we first estimate generalizability using mean $R$ scores and their standard deviations. There are three analysis categories: L, N-L, and N-L (best). For validation, we apply L and N-L models to data for four withheld subjects. Validation is not possible for N-L (best) since those models would depend on the withheld data, but an alternative approach for non-linear model selection was tested. Rather than selecting the top model during GP evolution, we selected the model which best generalized within-subject. For within-subject validation, these unbiased models are labelled N-L(UnB).

\subsection{Within-Subject Generalizability}

Within-subject generalizability is estimated by first applying models generated for one subject-run to data for the other, independent run, and then computing the $R$ score between the model output and the relevant hypothesized HR. 

\subsection{Between-Subject Generalizability}

Between-subject generalizability is estimated in two ways. First, given one of the two runs, we apply the models generated for one subject to all other subjects' data. We then compute the $R$ score between model outputs and the hypothesized HR. We call this approach \textit{pairwise between-subject generalizability}.

In a second approach, we use model output averaging to combine multiple models into one \textit{average model}. This approach simply applies multiple models to the same data and computes the average output. With 10 subjects, we generate 10 different average models, each with one subject left out. An average model is evaluated by applying it to the left-out subject's data and comparing the $R$ score between its output and the hypothesized HR. 

Validation was possible for between-subject generalizability due to there being four other subjects that were withheld from modelling. This was because of the computational cost for generating non-linear models. Pairwise between-subject generalizability is evaluated by applying generated models to withheld subjects' data and computing the $R$ score between the output and the relevant hypothesized HR. A similar validation strategy is taken for average models.

\subsection{Resting State Validation}

In an effort to compare how well linear and non-linear regression are able to fit task-free data to an unrelated hypothesized HR, we fit four models to a separate resting state scan. These consist of a linear and non-linear model for each of the two hypothesized HRs used in other experiments. This is especially important to do for non-linear regression since the number of degrees of freedom is so large. The $R$ score between the fitted model's output and the hypothesized HR provides an estimate of a modelling strategy's propensity to overfit. It would be insightful to have many more resting state scans available for analysis, but only one resting state scan was collected as part of the funded project. 

\section{Results}

\subsection{Within-Subject Generalizability}

\begin{figure}
	\centering
	\includegraphics[width=1.0\linewidth]{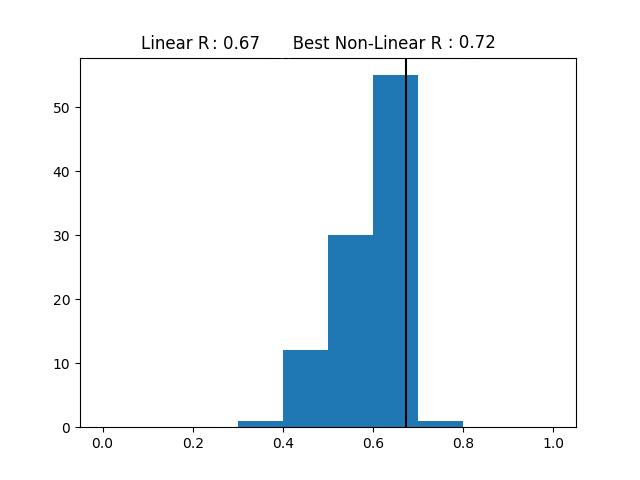}
	\caption{Distribution of non-linear model $R$ scores for within-subject generalizability. \textit{x-axis:} $R$ scores. \textit{y-axis:} Quantity of non-linear models in score interval. Example of the distribution of $R$ scores between non-linear models produced by GP and the hypothesized HR for 'Loc2' (\textit{all-stims}). The vertical black line denotes the score observed for the corresponding linear model. In this and most other cases, the number of non-linear models \textit{worse} than linear models forms a majority, but never a totality.}
	\label{fig:histogram}
\end{figure}

An estimation of the relative within-subject generalizability for linear and non-linear regression is summarized by Table \ref{table:within_sub}. We did not observe any significant difference between L and N-L models, however N-L (best) scores significantly higher than both L and N-L. It is important to note that all we can infer from this result is that GP produces, on average, at least one model per run that generalizes better than our linear models. A typical instance of this scenario is visualized by Figure \ref{fig:histogram}. Notice in this histogram that the majority of non-linear models generalized worse than their linear counterpart. However, in this case, and in fact in all other observed instances, there was at least one non-linear model with a better $R$ score than for the linear model.

Due to limitations on the number of models for which we could generate non-linear models, a separate validation set was not used for within-subject generalizability.

\begin{table}
	\centering{
	\begin{tabular}{|c|c|c|} 
		\hline 
		& \textit{all-stims} & \textit{faces} \\ 
		\hline 
		L & $\bar{x}=0.6035$ $s=0.1516$  &  $\bar{x}=0.2265$ $s=0.1229$\\ 
		\hline 
		N-L & $\bar{x}=0.5470$ $s=0.2151$  & $\bar{x}=0.1621$ $s=0.1522$ \\ 
		\hline 
		N-L (best) & $\bar{x}=0.7132$ $s=0.1302$ &  $\bar{x}=0.3240$ $s=0.1152$\\ 
		\hline 
	\end{tabular} 
	\caption{Mean $R$ scores ($\bar{x}$) and standard deviation ($s$) for \textbf{within-subject} generalizability. \textbf{\textit{all-stims}} In independent t-tests, L and N-L models are not significantly different, though N-L (best) scores significantly higher than L and N-L with $p < 0.022$ and $p < 0.006$, respectively. \textbf{\textit{faces:}} In independent t-tests, scores between L and N-L are not significantly different, though again N-L (best) scores significantly higher than both L and N-L with $p < 0.016$ and $p < 1\times 10^{-16}$, respectively.}
	\label{table:within_sub}
}
\end{table}

\subsection{Between-Subject Generalizability}
An estimation of the relative pairwise between-subject generalizability for linear and non-linear regression is summarized by Table \ref{table:between_sub_pair}. We observed in both categories that L models scored significantly higher than N-L models. And once again, we observed that N-L (best) generalizes better than both L and N-L. 

An estimation of average model generalizability is summarized by Table \ref{table:between_sub_average}. We observed improvements over pairwise between-subject generalizability in almost all cases. The only exception was for N-L (best) for \textit{faces}, which was not significantly different. Based on these results, we suggest that model averaging could be a useful technique for improving signal-to-noise ratios in between-subject modelling.

\begin{table} 
	\centering{
	\begin{tabular}{|c|c|c|} 
		\hline 
		& \textit{all-stims} & \textit{faces} \\ 
		\hline 
		L & $\bar{x}=0.4703$ $s=0.1959$  & $\bar{x}=0.1013$ $s=0.1182$ \\ 
		\hline 
		N-L & $\bar{x}=0.3864$ $s=0.2327$  & $\bar{x}=0.0754$ $s=0.1045$ \\ 
		\hline 
		N-L (best) & $\bar{x}=0.6399$ $s=0.1739$ & $\bar{x}=0.2814$ $s=0.1302$ \\ 
		\hline 
	\end{tabular} 
	\caption{Mean $R$ scores ($\bar{x}$) and standard deviation ($s$) for \textbf{pairwise between-subject} generalizability. \textbf{\textit{all-stims:}} In independent t-tests, L scores significantly higher than N-L with $p=0.2\times 10^{-3}$, N-L (best) scores significantly higher than N-L with $p < 1 \times 10^{-27}$, and N-L (best) scores significantly higher than L with $p=0.2\times 10^{-3}$. \textbf{\textit{faces:}} In independent t-tests, L scores significantly higher than N-L with $p<0.029$, N-L (best) scores significantly higher than N-L with $p < 1 \times 10^{-47}$, and N-L (best) scores significantly higher than L with $p < 1 \times 10^{-35}$. } 
	\label{table:between_sub_pair}
}
\end{table}

\begin{table}
	\centering{
	\begin{tabular}{|c|c|c|} 
		\hline 
		& \textit{all-stims} & \textit{faces} \\ 
		\hline 
		L & $\bar{x}=0.8183$ $s=0.0774$  & $\bar{x}=0.2389$ $s=0.0992$ \\ 
		\hline 
		N-L & $\bar{x}=0.7104$ $s=0.1889$  & $\bar{x}=0.1438$ $s=0.1293$ \\ 
		\hline 
		N-L (best) & $\bar{x}=0.8737$ $s=0.0620$ & $\bar{x}=0.2659$ $s=0.1410$ \\ 
		\hline 
	\end{tabular} 
	\caption{Mean $R$ scores ($\bar{x}$) and standard deviation ($s$) for \textbf{between-subject (average models)} generalizability. \textbf{\textit{all-stims:}} In independent t-tests, L scores higher than N-L with $p<0.027$ and N-L (best) scores higher than both L and N-L with $p<0.02$ and $<0.001$, respectively. \textbf{\textit{faces:}} In independent t-tests, none of the scores are significantly different from each other.} 
	 \label{table:between_sub_average}
	}
\end{table}

Pairwise between-subject generalizability validation results are summarized by Table \ref{table:pairwise_validation}. We compared these results to those in Table \ref{table:between_sub_pair}. According to independent t-tests, a significant difference between results for the modelled subjects and validation subjects was only observed for N-L, \textit{all-stims}, where the mean $R$ score was higher in validation with $p=0.034$.

\begin{table}
\centering{
\begin{tabular}{|c|c|c|}
	\hline 
	& \textit{all-stims} & \textit{faces} \\ 
	\hline 
	L & $\bar{x}=0.5287$ $s=0.1091$ & $\bar{x}=0.0550$ $s=0.1040$  \\ 
	\hline 
	N-L & $\bar{x}=0.4736$ $s=0.1637$  & $\bar{x}=0.0431$ $s=0.0832$  \\ 
	\hline 
	N-L(UnB) & $\bar{x}=0.5084$ $s=0.1821$  & $\bar{x}=0.0454$ $s=0.0831$ \\ 
	\hline 
\end{tabular} 
}
\caption{Mean $R$ scores ($\bar{x}$) and standard deviation ($s$) for \textbf{pairwise between-subject validation} on data from four withheld subjects. Within each stimulus category, there is no significant difference between modelling strategies.}
\label{table:pairwise_validation}
\end{table}

\begin{figure}
	\centering
	\includegraphics[width=1.0\linewidth]{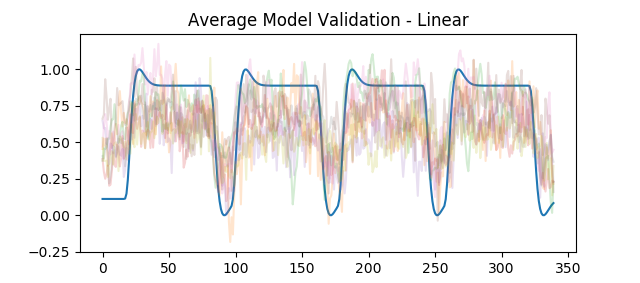}
	\includegraphics[width=1.0\linewidth]{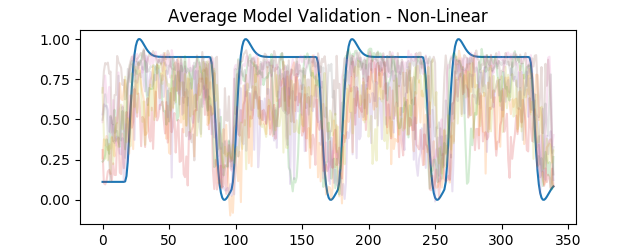}
	\includegraphics[width=1.0\linewidth]{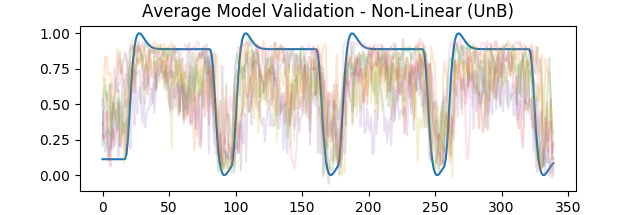}
	\caption{Average model validation for \textit{all-stims}. Superimposition of all models applied to validation subjects (translucent, many colours) against the hypothesized HR. Time series data are shown in arbitrary units over time in seconds. In validation, there was no significant difference between modelling strategies.}
	\label{fig:average_validation}
\end{figure}

Average model validation results are summarized by Table \ref{table:average_validation}. We compared these results to those in Table \ref{table:between_sub_average} and found that, in validation, generalizability tended to be worse than during model construction. Models for \textit{all-stims}, however, tend to be a good fit for the hypothesized HR, using any of the three modelling strategies. This result is visualized by Figure \ref{fig:average_validation}.

\begin{table}
	\centering{
		\begin{tabular}{|c|c|c|}
			\hline 
			& \textit{all-stims} & \textit{faces} \\ 
			\hline 
			L & $\bar{x}=0.6026$ $s=0.1288$ & $\bar{x}=0.1199$ $s=0.0477$  \\ 
			\hline 
			N-L & $\bar{x}=0.6491$ $s=0.1221$  & $\bar{x}=0.0893$ $s=0.0534$  \\ 
			\hline 
			N-L(UnB) & $\bar{x}=0.6591$ $s=0.1101$  & $\bar{x}=0.0670$ $s=0.0555$ \\ 
			\hline 
		\end{tabular} 
	}
	\caption{Mean $R$ scores ($\bar{x}$) and standard deviation ($s$) for \textbf{between-subject (average model) validation} on data from four withheld subjects. Within each stimulus category, there is no significant difference between modelling strategies.}
	\label{table:average_validation}
\end{table}

\subsection{Resting State Validation}
\begin{figure*}
	\centering
	\includegraphics[width=1.0\linewidth]{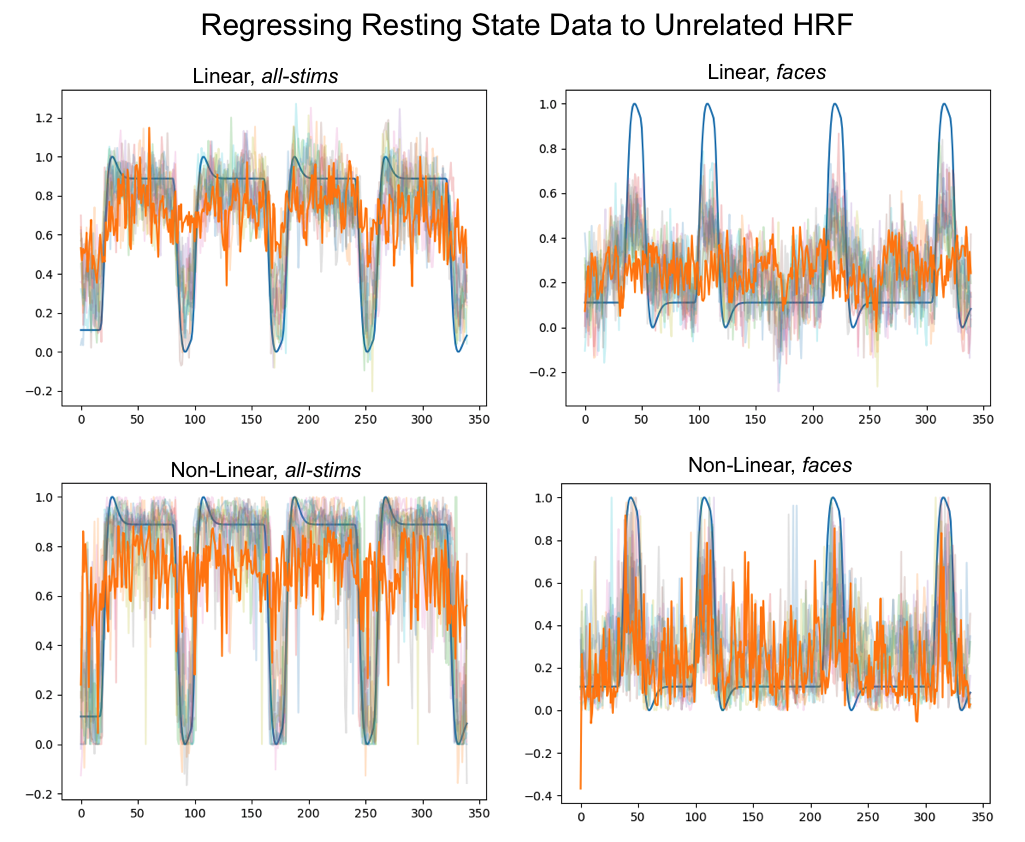}
	\caption{Visualization of outputs from models fitted using experimental data (translucent, multiple colours), the output from models fitted using unrelated resting state data (orange), and the stimulus-related hypothesized HR (blue). Associated $R$ scores are summarized by Table \ref{table:resting_valid}. Time series data are shown in arbitrary units over time in seconds.}
	\label{fig:restingsignals}
\end{figure*}

An estimation of each model's propensity to overfit, as estimated using independent resting state data, is summarized by Table \ref{table:resting_valid} and is visualized by Figure \ref{fig:restingsignals}. Though only one resting state scan is used, we can conclude that for both modelling strategies and for both stimulus categories, mean $R$ scores are significantly greater for stimulus-related models than for at least one resting state scan. This provides at least some confidence that models are not unreasonably overfit.

\begin{table}
	\centering{
\begin{tabular}{|c|c|c|}
	\hline 
	& \textit{all-stims} & \textit{faces} \\ 
	\hline 
	L & $\bar{x}_{exp}=0.83$  $R_{rest}=0.45$ & $\bar{x}_{exp}=0.59$  $R_{rest}=-0.08$  \\ 
	\hline 
	N-L & $\bar{x}_{exp}=0.86$  $R_{rest}=0.35$ & $\bar{x}_{exp}=0.56$  $R_{rest}=0.35$  \\ 
	\hline 
\end{tabular}
\caption{Mean $R$ scores for models fit using experimental data $(\bar{x}_{exp})$ compared to $R$ scores for models fit using unrelated \textbf{resting state} data $(R^2_{rest})$. In all cases, according to independent one-sample t-tests, resting state $R$ scores are significantly lower than mean experimental scores with $p < 1\times 10^{-6}$.}
\label{table:resting_valid}
}
\end{table}

\section{Discussion}

We found no evidence that non-linear regression is generally more prone to overfitting than linear regression. In resting state validation (see Table \ref{table:resting_valid}) linear regression produced a better-fit model than non-linear regression between completely independent time series. In practice, we can conclude only that both methods can be similarly prone to overfitting.

For within-subject generalizability, we observed that in all cases, at least one out of 100 GP runs produced a better model than linear regression. This is reflected by the consistently good results for N-L (best) models. The fact that such best-generalizing models exist does not help us select them in an unbiased manner, though. We were, however, able to use such results in an unbiased way by selecting the non-linear model that best generalized within-subject for between-subject analysis. Selecting such a model, however, did not yield \textit{significant} improvement over the other methods we tested.

Generalizability scores were poor for \textit{faces}. Though all models were significantly and positively correlated with the hypothesized HR, further work must be done to evaluate their specificity. The poor scores observed here could be due to using ROIs that are too large to capture face-selectivity. Even with the TalROI atlas, ROIs contained hundreds of voxels. A finer partitioning of face-selective areas or other atlases may yield better results. 

\section{Conclusions and Future Work}

The \textit{all-stims} hypothesized HR was very successfully modelled using either linear or non-linear regression. In experiments for both within- and between-subject generalizability, all models reproduced the hypothesized HR with mean $R$ scores reaching $0.65$ in validation, and up to $0.81$ in unbiased experiments. Furthermore, we found no evidence to suggest that non-linear regression over fit the data worse than linear regression. These results suggest that our linear and non-linear models generalized similarly well, and that neither was overfit more than the other.

In validation, we observed that model averaging significantly improves model quality for \textit{all-stims}, for all methods, when compared to pairwise between-subject generalizability. This suggests that all models feature a noise component that is partially subdued by averaging across subjects. This result is unsurprising, but provides further evidence that signal-to-noise ratios can be significantly improved by considering ROI-based averages across multiple subjects. 

Overall, we found that while they are not any more likely to over-fit, non-linear models did not generalize significantly better than linear models when unbiased model selection methods are applied. Though non-linear regression consistently produced \textit{some} model that is better than its linear counterpart, it is difficult to select and expensive to produce. As a result, we conclude that for small sample sizes, it seems far more sensible to use linear methods for multivariate regression modelling of fMRI data, pending improved methods for non-linear model selection.


\section{Acknowledgements}

 This research was supported in part by the Natural Sciences and Engineering Research Council of Canada (NSERC). 

 This research was enabled in part by support provided by Compute Ontario (\url{computeontario.ca}), Westgrid (\url{westgrid.ca}), and Compute Canada (\url{computecanada.ca}).

 Data collection costs were funded by a grant from the Canada First Research Excellence Fund (BrainsCAN: Brain Health for Life) to the University of Western Ontario.

 fMRI experimental design and stimuli were designed by Jody Culham and Rainer Goebel;  implemented by Kevin Stubbs.

\nocite{langley00}

\bibliography{JacksonHughesDaley}
\bibliographystyle{plainnat}

\begin{figure}
	\centering
	\includegraphics[width=1.0\linewidth]{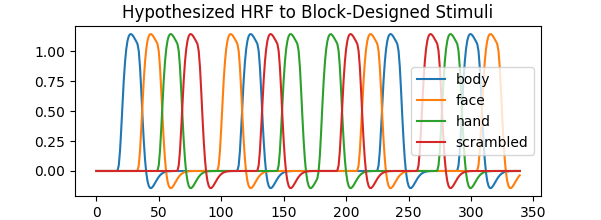}
	\caption{Hypothesized haemodynamic responses to stimuli in arbitrary units over time in seconds for `Loc1'. Stimuli (images) were presented to subjects for a constant duration of 0.8 seconds followed by a 0.2 second gap. The order of stimulus category presentation is reflected by the colours of the response curves. The other experimental run, 'Loc2' is a variation of 'Loc1' in which the order of stimulus category presentation was changed.}
	\label{fig:block}
\end{figure}

\newpage
\clearpage

\end{document}